\documentclass[aps,twocolumn,prl,tightenlines,floatfix,showpacs]{revtex4}
\usepackage[dvips]{graphicx}
\usepackage[english]{babel}
\usepackage{amsmath}
\usepackage{amssymb}
\usepackage{times}

\begin{document}

\title{Fermi Liquid Theory of Ultra-Cold Trapped Fermi Gases: Implications for
Pseudogap Physics and Other Strongly Correlated Phases}
\author{Chih-Chun Chien$^{1}$ and K. Levin$^{2}$}

\affiliation{$^1$Theoretical Division, Los Alamos National Laboratory, MS B213, Los Alamos, NM 87545, USA}
\affiliation{$^2$James Franck Institute and Department of Physics,
University of Chicago, Chicago, Illinois 60637, USA}

\date{\today}

\begin{abstract}

We show how Fermi liquid theory
can be applied to ultra-cold Fermi gases, thereby
expanding their ``simulation" capabilities to a class of
problems of interest to multiple physics sub-disciplines.
We introduce procedures for measuring and calculating 
position dependent Landau parameters. This lays the ground work for addressing
important controversial issues: (i) the
suggestion
that
thermodynamically, the normal
state of a unitary gas is indistinguishable from a Fermi liquid
(ii) that 
a fermionic system with strong repulsive contact interactions
is associated with either ferromagnetism or localization; this relates 
as well to $^3$He and its $p$-wave superfluidity.
\end{abstract}


\maketitle

Normal Fermi liquid theory describes low temperature fermionic quantum 
matter which is  
adiabatically connected to its non-interacting
counterpart
\cite{BaymPethick}. 
This theory, which has wide ranging implications for condensed 
and neutron star matter
as well as nuclei and nuclear physics \cite{NuclearFL}, 
revolves around a quantification
of Landau parameters. These are viewed as a collection of molecular field
contributions characterizing the 
interaction 
between renormalized
(quasi) particles. 
Because Fermi liquid theory is a major tool for addressing 
many-body systems and ultracold Fermi gases are 
a testing ground for counterpart theories, it is essential to establish
how to measure as well as calculate the associated Landau
parameters. This leads to the goal of the present paper: to
formulate the extension of Landau's important theory to an atomic
trapped gas.
Our work incorporates the local density approximation (LDA) which
is used in both our experimentally-based and theoretical analyses.
In the context of the latter, we present a methodology for calculating
the Landau parameters, given a microscopic
many body scheme.

This paper addresses the relationship between Fermi liquid theory and
two important situations: trapped gases in the
presence of moderately strong attractive as well as repulsive
contact interactions.
For the former the ground state  
is a superfluid in which
the
pairing strength
can be
continuously tuned from BCS to
Bose Einstein condensation (BEC) \cite{Ourreview}.
For intermediate strength  
attraction, associated with the ``unitary" phase,
of considerable
interest is
whether the normal state (above the transition
temperature $T_c$) is a Fermi liquid or not.
A body of evidence \cite{Grimm4,MITtomo,JinStrinati}
has arisen in support of the prediction \cite{JS2}
that somewhat above $T_c$, an unpolarized, unitary gas
contains pre-formed pairs associated with a
normal state gap or pseudogap.
\textit{If there is a gap in
the fermionic excitation spectrum, these unitary gases
would not qualify
as Fermi liquids}.
However, this has become a more controversial point,
recently when this observation was challenged \cite{SalomonFL}
through fits to
the measured temperature dependence of the pressure. 
The claim from Ref.~\cite{SalomonFL} that
``the low-temperature thermodynamics of the strongly interacting
normal phase is
well described by Fermi liquid theory", 
is both important and at odds with previous 
inferences \cite{Grimm4,MITtomo,JinStrinati,Ourreview,Chen4} that
Fermi liquid theory is inappropriate at unitarity. 

By contrast, for the case of repulsive interactions,
Fermi liquid theory should be widely applicable.
Nevertheless, there are on-going controversies about the
microscopic nature of such a Fermi liquid. In the cold gases this question
is posed in the context of considering  even stronger repulsion where
the central issue is
whether the ground state of the Hubbard gas model 
will be ferromagnetic \cite{KetterleFM,Ferrophase}
or localized \cite{ZhaiLocalized}.
Here we argue
that, in the Fermi liquid regime (at higher $T$, or weaker interactions so
that there is no spontaneous symmetry breaking)
one can use the behavior of the measured Fermi liquid parameters
as a
probe into the nature of the dominant incipient instability.
Indeed, this 
debate in the cold gases is completely analogous to a related controversy in the prototypical
Fermi liquid of condensed matter physics: $^3$He,  
which is thought to be similarly described by the repulsive Hubbard gas model. At
issue is whether the Landau parameters are controlled by proximity
to a ferromagnetic \cite{Valls} or localization \cite{VollhardtRMP}
instability.
Further insight into $^3$He should also be important to the cold gas
community in additional contexts, since the
$p$-wave
superfluidity of $^3$He
at $T=0$ can serve as
a template for addressing quite generally, non-$s$ wave
(\textit{e.g.,} $d$-wave) superfluidity which \textit{indirectly}
arises via repulsive on-site interactions.

An understanding of precisely what does and what does not constitute a
Fermi liquid is clearly lacking in the literature of ultracold Fermi gases,
where we have seen 
competing viewpoints expressed such as
between Ref.~\cite{SalomonFL} and Refs.~\cite{Grimm4,MITtomo,JinStrinati,Ourreview} or  
between Refs.~\cite{KetterleFM,Ferrophase} and Ref.~\cite{ZhaiLocalized},
or Refs.~\cite{Valls} and \cite{VollhardtRMP}.
There has been some discussion of the basis of 
Fermi-liquid theory, albeit limited
\cite{StringariFL}
to the dynamical response in
(less accessible) homogeneous systems.
Fermi liquid theory is thought to be relevant to strongly polarized unitary
gases, when superfluidity is destabilized. There are conjectures that a
rapidly rotating unitary gas (such that superfluidity is suppressed)
will also be a Fermi liquid \cite{StringariFL}, but this needs to be
systematically investigated since related analogous condensed matter systems show that pairing
survives even when superfluid coherence is destroyed \cite{Ourreview}.

\begin{figure}
  \includegraphics[width=3.in,clip] {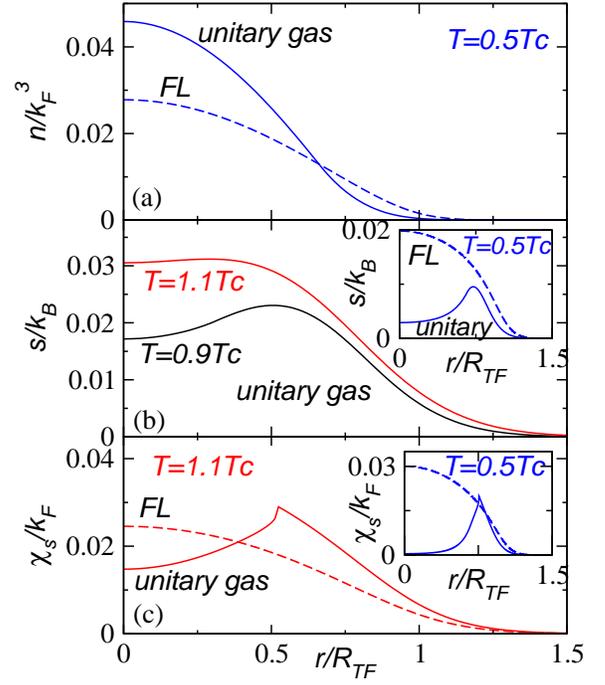} \caption{(Color
  online) Comparison of Fermi liquid (FL) and non-FL
  (unitary Fermi gas) thermodynamic behavior.  
In the FL case 
  we consider a repulsive contact interaction with
  $a=0.2a_{Stoner}$. (a) Density profile of a unitary Fermi gas at
  $T=0.5T_c$ (solid line) compared with that of the FL at the same $T$
  (dashed line). Here $T_c/T_F=0.27$.  
  (b) Entropy profiles close to $T_c$ of
  unitary Fermi gases, slightly above (red) and slightly below (black)
  $T_c$, and in the inset, at $T = 0.5T_c$, as compared with
  that of the FL
  (dashed line) for the same $T$.
(c) The spin susceptibility $\chi_s$ of a unitary Fermi gas
  (solid line, 
extracted from a
$5\%$ polarized gas with no phase separation) compared to that of the same Fermi liquid (dashed line) 
at $T=1.1T_c$. The inset
shows counterpart plots for $T=0.5T_c$.
Here $k_{F}(r=0) a_{Stoner}=\pi/2$. Pseudogap effects in (b),(c) are reflected
in the smooth evolution from above/below $T_c$ at unitarity.}
\label{fig:Ho_free}
\end{figure}

Because of these controversies and confusion, an important contribution of this paper is to
establish the signatures of Fermi liquid and non-Fermi
liquid behavior. In the process, we will also establish the thermodynamic
signatures of pseudogap effects. Figure~\ref{fig:Ho_free} 
is devoted to a comparison of a number of Fermi liquid 
properties with those of a non-Fermi liquid.
To represent the latter, we choose a unitary gas superfluid, for which there should
be no disagreement that normal Fermi liquid theory fails.
Figure~\ref{fig:Ho_free}(a)
presents plots of the calculated density profiles for the Fermi liquid (FL) and non-Fermi liquid
at the same temperature $T=0.5T_c$, where $T_c$ is the transition
temperature of the unitary superfluid.
For concreteness, in the FL case, we presume a
repulsive contact interaction of moderate size and use Hartree-Fock
theory (described below) to compute the profile.
As an experimental calibration,
for the unitary case both above and below $T_c$, the profiles are in
very good agreement \cite{JS5} with the data. Despite their evident
similarity,
the FL and non Fermi liquid density profiles contain very different physics
which must be carefully extracted.

These differences are best accessed
by studying derived physical properties. 
In this paper we focus on unpolarized Fermi gases. 
As a necessary first step, we compute
the compressibility $\kappa$, the spin
susceptibility and the entropy from
the density profiles. Within the LDA, the chemical potential is $\mu(r)=\mu(r=0)
-(1/2)m\omega_{tr}^{2}r^
{2}$, where $m$ is the bare
mass of the fermions and $\omega_{tr}$ is the trap
frequency. Then 
\begin{equation}\label{eq:dndmu}
\frac{dn}{d\mu}=-\frac{2}{m\omega_{tr}^{2}r}\frac{dn}{dr},
\end{equation}
which, in turn, allows one to measure
$\kappa\equiv
n^{-2}dn/d\mu$.
From the profiles of a slightly polarized gas
(with polarization $\le 5\%$)
one can arrive at the Pauli susceptibility $\chi_{s}\equiv\partial\delta
n/\partial\delta\mu\approx\delta n/\delta\mu$.
Here $\delta
n=n_{\uparrow}-n_{\downarrow}\ll n$ and $\delta
\mu=\mu_{\uparrow}-\mu_{\downarrow}$ and one requires
that
$\mu_{\sigma}$ and $n_{\sigma}$ must be established separately.
Measurements of $\chi_s$ have a firm basis in the past success of
LDA approaches to
polarized Fermi gases with attractive
interactions
\cite{Chienpolarized}.  
One requires
that the densities be small at the trap edge and that
(for attractive interactions) low polarizations at $T$
slightly away from zero are considered so that the phase separated state (with
discontinuous densities) is not encountered.

Techniques for extracting the entropy are directly based on those
discussed elsewhere \cite{HoNaturephys}, albeit modified somewhat.
We consider two clouds at nearby temperatures
$T_1$ and $T_2$ and label associated quantities of each
cloud with subscripts $1$ and $2$. 
Then by locating the points $\mu_{1}(T_1,r)=\mu_{2}(T_2,r^{\prime})$ 
on the two clouds \cite{HoNaturephys}
and using 
the Gibbs-Duhem relation, 
\begin{eqnarray}\label{eq:dPdT}
\left(\frac{\Delta P}{\Delta T}\right)_{\mu} 
&=& s(T_2)+\frac{1}{2}\left(\frac{ds(T_2)}{dT}\right)_{\mu}\Delta T ~~\mbox{  (in general)} \nonumber \\
&=&s\left(\frac{T_1+T_2}{2}\right) ~~\mbox{  (Fermi liquids).}
\end{eqnarray}
We note that 
the extracted $s(r)$ agrees very
well with the exact $s(r)$ provided one uses the \textit{averaged} $T$. 
With care, one should be able to choose the separation between $T_1$ and $T_2$ large enough 
so that noise effects do not affect the extraction of $s(r)$.

We show in the inset to Fig.~\ref{fig:Ho_free}(b) a plot of $s(r)$
calculated \cite{ChenThermo} at unitarity as compared with the FL case, for
$T/T_c = 0.5$. 
In the main body of
Fig.~\ref{fig:Ho_free}(b) we compare this with the counterpart unitary plots just
below the transition temperature ($T/T_c = 0.9$) and in the normal
phase at $T/T_c = 1.1$. The total entropy involves a bosonic as well
as a fermionic contribution \cite{ChenThermo}. 
A salient feature of $s(r)$ in these unitary superfluids is that it is
spatially non-monotonic. 
This behavior is associated with
a pairing gap which is largest
at the trap center and decreasing with increasing $r$.
That this non-monotonicity in $s(r)$ is found at $T/T_c =
0.9$ suggests the pairing gap is still large at these high temperatures; 
because
the transition is second order this pairing gap must persist into the normal phase,
as a 
``pseudogap" \cite{Ourreview,ourNSR,Chen4}
associated with non-condensed pairs. Indeed, one sees that the curves for
$T/T_c = 1.1$
exhibit an (albeit, slightly) anomalous non-monotonic entropy profile.
%

Similar non-monotonic behavior can also be found in the
spin susceptibility $\chi_{s}$ of an unpolarized unitary Fermi gas which is
plotted in Fig.~\ref{fig:Ho_free}(c) for $T/T_c=1.1$ 
(and in the inset at $T/T_c = 0.5$)
compared to that of the same Fermi liquid
as discussed in
Fig.~\ref{fig:Ho_free}(a).
These two non-monotonicities contrast
with their counterparts for the Fermi liquid case shown in the figures.
The suppression of $\chi_s$ at the trap center of the normal unitary
gas can similarly be traced to
singlet non-condensed pairs
\cite{Chienpolarized}, or pseudogap effects.

\textit{The behavior of the normal unitary gas, thus, shows a continuous evolution
from the superfluid properties (say, measured at $T/T_c \approx 0.5$) to
above $T_c$. This is the sense in which pseudogap
effects (manifestly inconsistent with Fermi liquid theory)
will be evident in thermodynamics}. This is in contrast
to the claim, based on a quadratic temperature
dependence of the pressure \cite{SalomonFL}
that ``the normal phase of the unitary Fermi
gas.. [has] thermodynamic properties [which are] .. well described
by Fermi liquid theory, unlike high-$T_c$ copper oxides." 
The same quadratic
signature
\cite{SalomonFL} can also
be found theoretically
for a range of $T$ in strict BCS theory, where FL theory 
obviously fails.
Thus, we infer that power law dependences in thermodynamical
variables 
are not valid indicators of the applicability of Fermi liquid theory.
Rather, non-monotonicity in $s(r)$ and $\chi_{s}$ profiles, which reflects
the $T < T_c$ physics, is 
the more relevant feature, a consequence 
but not a proof of the existence of a pseudogap 
\cite{Grimm4,MITtomo,JinStrinati}
and the concomitant
failure of Fermi liquid theory. 
Interestingly, a number of the theories with which these
experiments \cite{SalomonFL} are
favorably compared also contain pseudogap effects.

In
proper Fermi liquids
the compressibility, spin susceptibility and entropy enable one to
experimentally extract the spatially dependent Landau parameters
\cite{BaymPethick} and effective mass $m^*$.
In condensed matter Fermi
liquids $m^*$ is most readily deduced from
the linear (in temperature) term in the specific heat \cite{BaymPethick}.
In cold gases, we argue that it
is best obtained from
the linear in temperature term in
the local entropy density
$s(r)$:
\begin{equation}\label{eq:s_and_mstar}
s(r)=\frac{m^*k_F(r)T}{3}.
\end{equation}
Here we set $\hbar=1$ and $k_{B}=1$. Fermi liquid theory and Eq.~(\ref{eq:s_and_mstar}) are precise
only if $T\ll T_F$. As the local $T_F(r)$
decreases, this procedure will fail towards
the trap edge.
Thus expressions including only leading order $T$ corrections in $s(r)$ (or
equivalently
$T^{2}$ terms in $P$ or $\mu$, etc.) must be constrained to
$T\ll T_F(r)$. 
Moreover, in
a Galilean invariant system \cite{BaymPethick}, 
$F_{1}^{s}$ satisfies
\begin{equation}\label{eq:F1s}
\frac{m^{*}(r)}{m}=1+\frac{F_{1}^{s}(r)}{3}.
\end{equation}
and these arguments apply to a trapped gas within the LDA.
%
The Landau parameter $F_{0}^{s}$ can be obtained from 
\begin{equation}\label{eq:F0s}
\frac{dn}{d\mu}=\left(\frac{dn_0}{d\mu}\right)\frac{m^{*}/m}{1+F_{0}^{s}}.
\end{equation}
Here $dn_0/d\mu=-\sum_{\bf k}(\partial f(\omega)/\partial\omega)|_{\omega=\hbar^{2}k^{2}/2m-\mu}$ is the known
corresponding quantity for a non-interacting
Fermi gas at the same chemical potential and $f(x)$ is the Fermi distribution function.
Given $dn/d\mu(r)$ and
$m^{*}(r)/m$, $F_{0}^{s}(r)$ is then known.
Finally, the
third Landau parameter to be quantified
is $F_{0}^{a}$, which
is related to the Pauli susceptibility 
\begin{equation}\label{eq:F0a}
\chi_{s}=\chi_{s0}\frac{m^{*}/m}{1+F_{0}^{a}}.
\end{equation}
Here $\chi_{s0}
=d\delta n_0/d\delta\mu$
is the spin susceptibility of the 
non-interacting Fermi gas.

We consider now a repulsive contact interaction
in order to gain more understanding of the
stable phases of the Hubbard Fermi gas Hamiltonian.
This bears on recent 
\cite{KetterleFM,ZhaiLocalized}
and longstanding controversies
\cite{Valls,VollhardtRMP} in the literature.  If
the repulsion is not too large, one expects the system to be in a
Fermi liquid phase.  Our methodology for computing the associated
Landau parameters is based on first computing the trap profiles
$n_{\sigma}(r)$ from a microscopic theory and secondly following
the experimental protocols outlined above to determine
the LDA-derived $F_{1}^{s}$, $F_{0}^{s}$,
and $F_{0}^{a}$. The specifics of a microscopic theory are tested by
comparing
the calculated Landau parameters with
their experimental counterparts.
In LDA the density is 
$n_{\sigma}(r)=\sum_{{\bf k},\omega_n} G_{\sigma}(\Sigma_{\sigma}({\bf k},
\omega_n, \mu(r)) $.  Here $G_{\sigma}$ is the Green's function, $\omega_n$ is the fermion Matsubara
frequency, and the fermion
self-energy $\Sigma_{\sigma} ({\bf k}, \omega_n, \mu(r))$ is chosen
to 
correspond, 
for example, to a nearly ferromagnetic \cite{Valls}
or nearly
localized \cite{VollhardtRMP} theory.

While it is relatively straightforward to use more
sophisticated approaches \cite{Valls}, in order to focus on
the central physics, we apply Hartree-Fock theory
\cite{Ferrophase,TosiHartree}. 
Here
$\Sigma_{\sigma}=gn_{\bar{\sigma}}$, where $g=4\pi\hbar^{2}a/m$,
$\bar{\sigma}=-\sigma$, and $a$ is the two-body $s$-wave scattering
length. The trap profiles correspond to $n_{\sigma}(r)=\sum_{\bf
k}f\left(\frac{\hbar^{2}k^{2}}{2m}+gn_{\bar{\sigma}}(r)-\mu_{\sigma}(r)\right)$.
We consider
$a/a_{Stoner}=0.6$ and $0.9$ at $T/T_F=0.02$ and $0.01$,
where $a_{Stoner}$ is the critical scattering length (Stoner
instability) at which
$\chi_{s}$ diverges. 
\begin{figure}
  \includegraphics[width=3.in,clip] 
{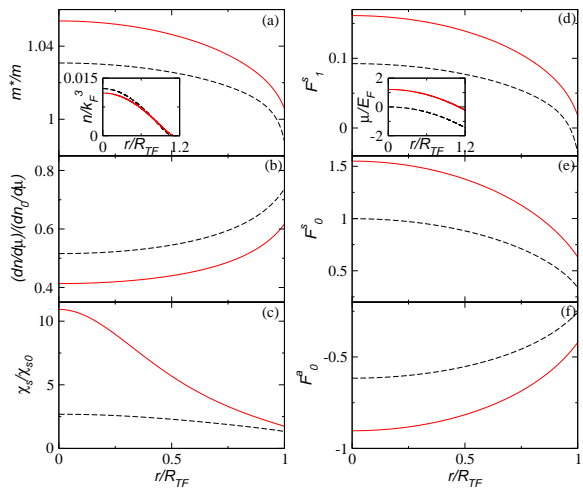}
  \caption{(Color online) The extracted (a) $m^*/m$, (b)
  $(dn/d\mu)/(dn_0/d\mu)$, and (c) $\chi_{s}/\chi_{s0}$ and the
  Landau parameters (d) $F_{1}^{s}$, (e) $F_{0}^{s}$, and (f)
  $F_{0}^{a}$. The input data are from the Hartree-Fock theory with
  $a/a_{Stoner}=0.6$ (black dashed line) and $0.9$ (red solid line) at $T/T_F=0.02$ (whose
  $n(r)$ and $\mu(r)$ are shown as insets) and $0.01$.}
\label{fig:Landau_param}
\end{figure} 
From the profiles, one obtains
$m^{*}$, $(dn/d\mu)$, and $\chi_{s}$. The corresponding
spatial dependences associated with each of these
are shown on the left column of
Fig.~\ref{fig:Landau_param}, where we 
focus on the results at the trap center.
Using Eqs.~(\ref{eq:F1s}), (\ref{eq:F0s}), and (\ref{eq:F0a}), the
three Landau parameters $F_{1}^{s}$, $F_{0}^{s}$, and $F_{0}^{a}$ are
obtained and plotted in the right column of
Fig.~\ref{fig:Landau_param}. We see that $m^{*}/m$ is only
slightly larger than $1$ so $F_{1}^{s}$ is small. 
For repulsive interactions, $F_{0}^{s}$
is expected to be positive and
$F_{0}^{a}$ is expected to be negative, as is consistent with
Fig.~\ref{fig:Landau_param} panels (e) and (f). 

Interestingly, had one naively applied these same FL procedures to the
attractive case (where we argue because of the pseudogap,
 Fermi liquid theory
is not appropriate) 
one would
find that 
$F_{0}^{a}$
is positive and
$F_{0}^{s}$ is negative.
Therefore, even though the trap profiles for the particle
density are not qualitatively
different for both cases, the derived Landau parameters make it
possible to distinguish between very different physical
situations.

One can see from Fig.~\ref{fig:Landau_param}(d) and (f)
that as 
$a\rightarrow a_{Stoner}$
$F^{a}_{0}\rightarrow -1$
while $F^{s}_{1}$ and the effective mass remain finite.
In contrast,
for an almost localized liquid
\cite{ZhaiLocalized}, 
$F^{s}_{1}$
or $m^*/m$ will diverge at the counterpart critical interaction strength.
While
this leads to a divergence in $\chi_{s}$,
$F^{a}_{0}$ does not approach $-1$ \cite{VollhardtRMP}. 
For
an unknown ground state, this Fermi liquid analysis in the
limit that the interaction strength is slightly reduced, can be used to
provide important hints about the nature of
the more strongly
correlated phase.

Thus, future measurements of the Landau parameters associated
with the repulsive Hubbard gas studied in Ref.~
\onlinecite{KetterleFM} should be able to shed light on
a current controversy \cite{ZhaiLocalized} over whether a
ferromagnetic ground state has been observed.
There are, however, some
complications in these experiments.
The
Feshbach
resonance which tunes the repulsive interactions between
fermions is associated with 
a molecular bound state lying below the particle
continuum which is populated \cite{KetterleFM} at the $25\%$ level.
These molecules derive from molecule-particle or
molecule-molecule scattering which may lead to systematic
errors in determining $dn/d\mu$ and
$\chi_{s}$. 
To resolve this issue, one has to either
reduce the population of molecules or develop a more
sophisticated theory to take them into account.

An important contribution of this paper has been to establish
how, despite previous claims \cite{SalomonFL}, 
pseudogap effects do enter into thermodynamical measurements.
When a pseudogap is present, thermodynamical variables above
$T_c$ will show a smooth evolution from their behavior deep in
the superfluid phase;
this contrasts with BCS theory, where
the normal state thermodynamics do not
reflect their superfluid counterparts.
While there will be a thermodynamical
feature of the phase transition \cite{Chen4}
one should see this smooth evolution in anomalous \textit{spatial dependences}
for example of the spin susceptibility.
The experimental complexity needed to search for these pseudogap
effects
is comparable to
that in
Refs.~\cite{SalomonFL,Mukaiyama}. 
In this paper, following
Ref.~\onlinecite{KetterleFM}
we have also mapped out a new direction for
the field of ultra-cold trapped Fermi gases which, via 
simulations of the (repulsive) Fermi Hubbard gas, can serve to
establish where ferromagnetism,
localization and $p$ wave superfluidity are stable.
This should be facilitated by analyzing simultaneous 
measurements and calculations
of appropriate Landau parameters.

This work was supported by Grant No. NSF-MRSEC
DMR-0213745. We thank Q. J. Chen for providing thermodynamical plots.
C.C.C. acknowledges the support of the U.S.
Department of Energy through the LANL/LDRD Program.

\bibliographystyle{apsrev}

\end{document}